\documentclass[sigconf]{acmart}%

\AtBeginDocument{%
  \providecommand\BibTeX{{%
    \normalfont B\kern-0.5em{\scshape i\kern-0.25em b}\kern-0.8em\TeX}}}

\usepackage{amsmath,amsfonts}
\usepackage{mathtools}
\usepackage{enumitem}
\usepackage[algoruled, noend]{algorithm2e}

\usepackage{url}
\usepackage{booktabs}
\usepackage{bigdelim}
\newtheorem{problem}{Problem}
\usepackage{hyperref}

\newcommand{\introbox}[1]{%
    \par\addvspace{\smallskipamount}%
    \noindent\hspace{0.02\linewidth}%
    \fbox{
    \begin{minipage}{0.92\linewidth}%
    \textbf{#1}%
    \end{minipage}%
    }%
    \par\addvspace{\smallskipamount}%
}

\newcommand{\surf}{{SuRF}\xspace}
\newcommand{\rosetta}{{Rosetta}\xspace}
\newcommand{\snarf}{{SNARF}\xspace}
\newcommand{\proteus}{{Proteus}\xspace}
\newcommand{\bloomrf}{{bloomRF}\xspace}
\newcommand{\rencoder}{{REncoder}\xspace}
\newcommand{\rencoderse}{{REncoderSE}\xspace}
\newcommand{\rencoderss}{{REncoderSS}\xspace}

\newcommand{\budget}{\ensuremath{B}}
\newcommand{\bv}{\ensuremath{C}}
\newcommand{\sel}{\ensuremath{\mathit{select}}}

\newcommand{\bzero}{\texttt{0}\xspace}
\newcommand{\bone}{\texttt{1}\xspace}
\newcommand{\lo}[1]{\ensuremath{\cramped{#1^\mathit{lo}}}}
\newcommand{\hi}[1]{\ensuremath{\cramped{#1^\mathit{hi}}}}

\newcommand{\myparagraph}[1]{\medbreak\noindent\textbf{#1.}}

\usepackage{tikz}
\newcommand*\circled[1]{\tikz[baseline=(char.base)]{
            \node[shape=circle,fill,inner sep=1pt] (char) {\textcolor{white}{#1}};}}

\begin{document}

\title{Grafite: Taming Adversarial Queries with Optimal Range Filters}

\author{Marco Costa}
\affiliation{%
  \institution{University of Pisa}
  \country{Italy}
}
\email{m.costa22@studenti.unipi.it}

\author{Paolo Ferragina}
\orcid{0000-0003-1353-360X}
\affiliation{%
  \institution{University of Pisa}
  \country{Italy}
}
\email{paolo.ferragina@unipi.it}

\author{Giorgio Vinciguerra}
\orcid{0000-0003-0328-7791}
\affiliation{%
  \institution{University of Pisa}
  \country{Italy}
}
\email{giorgio.vinciguerra@unipi.it}

\begin{abstract}
Range filters allow checking whether a query range intersects a given set of keys with a chance of returning a false positive answer, thus generalising the functionality of Bloom filters from point to range queries.
Existing practical range filters have addressed this problem heuristically, resulting in high false positive rates and query times when dealing with adversarial inputs, such as in the common scenario where queries are correlated with the keys.

\looseness=-1
We introduce Grafite, a novel range filter that solves these issues with a simple design and clear theoretical guarantees that hold regardless of the input data and query distribution: given a fixed space budget of $\budget$ bits per key, the query time is $O(1)$, and the false positive probability is upper bounded by $\ell/2^{\budget-2}$, where $\ell$ is the query range size.
Our experimental evaluation shows that Grafite is the only range filter to date to achieve robust and predictable false positive rates across all combinations of datasets, query workloads, and range sizes, while providing faster queries and construction times, and dominating all competitors in the case of correlated queries.

As a further contribution, we introduce a very simple heuristic range filter whose performance on uncorrelated queries is very close to or better than the one achieved by the best heuristic range filters proposed in the literature so~far.
\end{abstract}

\ccsdesc[500]{Theory of computation~Bloom filters and hashing}
\ccsdesc[500]{Information systems~Unidimensional range search}

\keywords{range filter, Bloom filter, range search, data structure}

\setcopyright{none}
\settopmatter{printacmref=false} %
\renewcommand\footnotetextcopyrightpermission[1]{} %
\pagestyle{plain}
\titlenote{This work has been accepted for publication in Proceedings of the ACM on Management of Data (SIGMOD 2024).}

\maketitle

\section{Introduction}\label{sec:intro}

Filters are data structures that allow checking whether a query key belongs to a given set of keys, with a chance of returning a false positive answer in exchange for a small space occupancy, i.e. much smaller than the storage of the full set.

Due to their compactness and the guarantee of not returning false negatives, filters are often kept in main memory and used to prevent unnecessary and costly accesses and searches in the set.
For example, they can avoid unnecessary network communications if a remote server does not contain the sought resource, or they can avoid unnecessary disk reads when the set is stored on disk.
In fact, since their introduction by Bloom~\cite{Bloom:1979} in 1970s, filters have been successfully used in networking~\cite{broderNetworkApplicationsBloom2004}, distributed systems~\cite{Tarkoma:2012}, databases~\cite{Dayan:2018}, bioinformatics tools~\cite{Chikhi:2021}, and search engines~\cite{Goodwin:2017}, to mention just a few applications.

While the vast majority of filters are capable of answering approximate membership (point) queries~\cite{Luo:2019,Fan:2014,Graf:2020,dillinger2022burr,Porat:2009,paghOptimalBloomFilter2005,Dayan2023infinifilter}, a new line of research, started a decade ago~\cite{alexiou2013adaptive}, focused on their generalisation to \emph{range queries}, which occur frequently in big data systems such as key-value stores~\cite{goswamiApproximateRangeEmptiness2014,zhangSurfPracticalRange2018,luoRosettaRobustSpacetime2020,vaidyaSNARFLearningenhancedRange2022,knorr2022proteus}. In this case, the filtering problem can be formally stated as follows.

\begin{figure}
    \centering
    \includegraphics{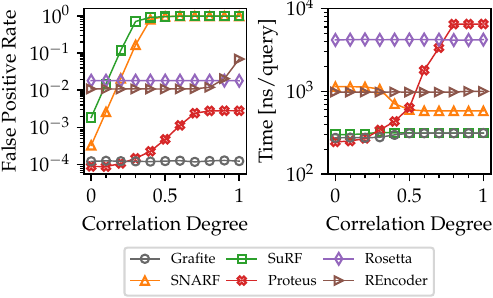}
    \caption{Grafite is the only range filter to date that is both effective (low false positive rate) and efficient (low query time) as the endpoints of the query range get closer to the~data.}
    \label{fig:motivation}
\end{figure}

\begin{problem}[Approximate range emptiness]
Given a set~$S$ of $n$ keys drawn from an integer universe $[u] = \{0, \ldots, u - 1 \}$, build a space-efficient data structure, called \emph{range filter}, that answers {\em range emptiness queries} of the form $[a, b] \cap S  \neq \emptyset\,?$ for any  $a\leq b$ in $[u]$. The range filter is allowed to return ``not empty'' when actually $[a, b] \cap S = \emptyset$ (i.e. a false-positive error) with probability at most~$\varepsilon$.
\end{problem}

From a theoretical point of view, this problem was solved optimally by Goswami~et~al.~\cite{goswamiApproximateRangeEmptiness2014}, which first proved a space lower bound of $\Omega(\log\tfrac{L}{\varepsilon})-O(1)$ bits per key, where $L$ is an upper bound on the query range size, and then they gave a data structure for the $w$-bit word RAM model matching this space up to a lower order additive term and offering constant-time queries when $w =\Omega(\log \tfrac{nL}{\varepsilon})$.

From a practical point of view, the literature offers a vast choice of range filters, such as ARF~\cite{alexiou2013adaptive}, \surf~\cite{zhangSurfPracticalRange2018}, \rosetta~\cite{luoRosettaRobustSpacetime2020}, \snarf~\cite{vaidyaSNARFLearningenhancedRange2022}, \proteus~\cite{knorr2022proteus},  \bloomrf~\cite{mossner2022bloomrf}, and \rencoder~\cite{Wang2023rencoder}.
These solutions, reviewed in Section~\ref{sec:related}, adopt totally different approaches to range filtering, thus offering a large number of trade-offs among space, \emph{empirical} probability of a false positive error (henceforth, false positive rate), query time, and construction time.
This notwithstanding, there is still one fundamental challenge that the literature has not yet been able to address:

\introbox{No practical solution is robust enough to efficiently handle all input data and query distributions.}

\noindent
Existing practical range filters, indeed, adopt heuristic designs that sacrifice performance guarantees to improve upon some specific inputs.
In fact, these range filters hardly guarantee a bounded false positive probability $\varepsilon$ for a given amount of space, thus, strictly speaking, they do not solve the approximate range emptiness problem unless some specific (and strong) assumptions on the kind of query workload and input data distribution are met.

As a consequence of this, there exist \emph{adversarial distributions} that can drive the false positive rate arbitrarily close to 1, thus making the filter useless, if not dangerous for the big data systems making use of it (e.g. because of increased disk or network activity that the filter was actually deployed to prevent).
The importance of this issue has also been stressed by Knorr et al.~\cite{knorr2022proteus}, who after presenting a formal framework of existing range filters, conclude that {\em ``no current design can handle [adversarial workloads] practically, suggesting the need for further expansion of the range filter design~space.''}

Notably, the vast majority of range filters suffer from the so-called \emph{correlation} between keys and queries, that is, they provide little or no filtering at all when an endpoint of the query range is close to one of the keys in the input set, which is quite disappointing given the commonness of such a workload in applications that care about the local properties of data (such as in time series applications where we need to check if some events occurred in a time frame)~\cite{luoRosettaRobustSpacetime2020}, or given that malicious users can artificially issue these queries with just the knowledge of (a subset of) the keys.
To demonstrate this issue, we show in Figure~\ref{fig:motivation} how existing range filters quickly reach high false positive rates as the endpoints of the query range get closer to the input keys, denoted as ``correlation degree'' on the horizontal axis (and detailed in our experimental section).
This holds true for \surf, \snarf, \rencoder, and \proteus, the latter even being auto-tuned on (i.e. overfitted to) the query workload.
The only exception is Rosetta which has a constant false positive rate but a query time that is up to orders of magnitude higher compared to the other filters.

\noindent Apart from the lack of robustness, there is another challenge:

\introbox{Current range filters are complex to evaluate and deploy because of their complicated design.}

\noindent
As stated above, existing range filters adopt complex design choices aimed at increasing their efficacy on some specific inputs.
For instance, SuRF~\cite{zhangSurfPracticalRange2018} encodes a trie with input keys truncated at their  distinguishing prefix (thus providing better filtering when there is no correlation), while SNARF~\cite{vaidyaSNARFLearningenhancedRange2022} maps each input key to a $\bone$-bit in a bitvector via a model learned from the data (thus providing better filtering when there are no outliers or poisoned data~\cite{Kornaropoulos:2022}).

These designs, coupled with the lack of guaranteed bounds on the false positive rate, hinder our understanding of how the range filter will behave once deployed to production unless future data and queries will follow exactly the same distribution of the test data on which the \emph{empirical} false positive rate was originally observed.
The ability to auto-tune on a sample of queries and input keys, as in Proteus~\cite{knorr2022proteus},  only partially eases the hard job of integrating a range filter into a real system, as there is still the necessity to keep a proper set of sample queries (thus also allocating further space) and to detect when the filter needs to be rebuilt because of workload shifts (thus introducing additional delays and requiring to keep the input data in memories close to where the range filter is built).

Instead, we aspire to a practical range filter that, similarly to Bloom filters, works robustly out of the box regardless of the input data and future queries, while hiding the complexities of its design and exposing just simple knobs such as the false positive probability~$\varepsilon$ or a space budget.

\myparagraph{Our contributions}
\begin{itemize}[leftmargin=*]
  \item We introduce \emph{Grafite}, a novel practical range filter that solves the lack of robustness and the high complexity of current solutions. Unlike all the practical range filters to date, Grafite offers clear guarantees that hold regardless of the input data and query distributions: given a fixed space budget of $\budget$ bits per key, the query time is $O(1)$, and the false positive probability is upper bounded by $\min\{1,\ell/2^{\budget-2}\}$, where $\ell$ is the query range size. Perhaps surprisingly, this is achieved via a simple design that maps the input keys into a smaller universe via a properly designed hash function~\cite{goswamiApproximateRangeEmptiness2014}, stores the resulting hash codes space-efficiently~\cite{eliasEfficientStorageRetrieval1974,fanoNumberBitsRequired1971}, and checks hash codes for inclusion in a range via an efficient query algorithm.

  \item We provide a comprehensive related work section and propose the first theoretical comparison of the space-time performance of range filters, showing the superiority of Grafite over prior solutions.

  \item We perform the largest experimental comparison among range filters, both in terms of dataset size and in the number of tested solutions, which shows that \emph{all} the existing filters provide little to no filtering or a high query time in the case of correlated query workloads.
  Instead, Grafite is the only range filter to date to achieve a robust and predictable false positive rate across all combinations of datasets, query workloads, and range sizes, while also providing faster queries and construction times, and dominating all competitors in the case of correlated query workloads.

  \item For datasets and uncorrelated query workloads previously tested in the literature, we show that there exists a very simple heuristic filter design\,---\,that we name \emph{Bucketing}\,---\,that essentially matches the filtering effectiveness of all the existing heuristic range filters, which are however significantly more complex and incur in higher query and construction times. This demonstrates that, if we give up on robustness guarantees, the approximate range emptiness problem can sometimes be addressed with a very simple solution.
\end{itemize}

\myparagraph{Paper outline}
Section~\ref{sec:related} discusses existing range filters.
Section~\ref{sec:grafite} introduces Grafite.
Section~\ref{sec:bucketing} introduces Bucketing.
Section~\ref{sec:theoretical} compares the space-time bounds of Grafite with those of existing range filters. 
Section~\ref{sec:experiments} experiments with Grafite, Bucketing and existing range filters.
Section~\ref{sec:conclusion} concludes the paper and suggests some open problems.

\section{Related Work}
\label{sec:related}

Consider an upper bound $L$ on the query range size $b - a + 1$. We can provide a trivial solution to the approximate range emptiness problem by using point filters which, given a false positive probability of $\gamma$, can be implemented in $n\log\tfrac{1}{\gamma} + O(n)$~bits of space and $O(1)$ query time~\cite{paghOptimalBloomFilter2005,Porat:2009}.
Indeed, by building a point filter on the input set $S$ with false positive probability $\gamma = \varepsilon/L$, we can check the existence of any element of $[a,b]$ in $S$ by executing at most $L$ point queries. This solution takes $n\log\tfrac{L}{\varepsilon} + O(n)$ bits of space, $O(L)$~query time, and the false positive probability is at most~$\varepsilon$ by union bound.

The question is now how far is this trivial solution from being optimal. The answer was given by Goswami~et~al.~\cite{goswamiApproximateRangeEmptiness2014}, which proved the following lower bound.

\begin{theorem}[\cite{goswamiApproximateRangeEmptiness2014}]
    \label{thm:lowerbound}
    Any data structure solving approximate range emptiness queries of fixed length $L \leq u/(5n)$ on $n$ keys drawn from an integer universe $[u] = \{0, \ldots, u - 1 \}$ with a false positive probability of $\varepsilon$ must use at least $n \log\big(\tfrac{L^{1-O(\varepsilon)}}{\varepsilon}\big) - O(n)$ bits of space.
\end{theorem}

This is a disappointing result because it states that, for a sufficiently small $\varepsilon$, at least $\log\tfrac{L}{\varepsilon}$ bits per key are needed.
Hence, the larger is $L$ and/or the smaller is $\varepsilon$, the larger is the space required by any range filter.
Note that we can restrict $L \leq u\varepsilon/n$, since otherwise it is more convenient to store the input keys in space close to $\log\tfrac{u}{n}$~bits per key (e.g. with an Elias-Fano encoding~\cite{eliasEfficientStorageRetrieval1974,fanoNumberBitsRequired1971}) thus solving the problem without false positives (i.e. $\varepsilon=0$).

Furthermore, Theorem~\ref{thm:lowerbound} implies that we cannot improve the space occupancy of the trivial solution stated above, but it challenges us to find a solution that matches its same space bound whilst improving the unattractive $O(L)$ query time. In this respect, Goswami~et~al.~\cite{goswamiApproximateRangeEmptiness2014} also introduce a  data structure that solves the range emptiness problem in $n \log\tfrac{L}{\varepsilon} + o{(n\log\tfrac{L}{\varepsilon})} + O(n)$~bits of space, while offering $O((\log\tfrac{nL}{\varepsilon})/w)$ query time in the $w$-bit word RAM model, thus achieving $O(1)$ query time when $w =\Omega(\log \tfrac{nL}{\varepsilon})$. 
The overall approach is mainly theoretic in nature and thus very complicated to implement. Nevertheless, the idea in~\cite{goswamiApproximateRangeEmptiness2014}  to reduce the original universe $U$ into a smaller universe $h(U)$ via a proper hash function $h$ is effective, and it will be used in Grafite too.

We now turn our attention to practical range filters.

\myparagraph{Prefix Bloom Filter} 
A Prefix Bloom Filter hashes key prefixes of a predetermined bit-length $l$ within a Bloom filter~\cite{dharmapurikarLongestPrefixMatching2003,matsunobu2020myrocks}.
Since each prefix encodes a range of the universe of size $2^{\log u - l}$, the filter can answer a range emptiness query by probing each range (i.e. configuration of $l$ bits) that overlaps with the query range, and returning ``empty'' if all the probes return false, ``not empty'' otherwise.
We do not further consider Prefix Bloom Filters because they are generalised by Rosetta~\cite{luoRosettaRobustSpacetime2020} and Proteus~\cite{knorr2022proteus}, which are described below and used in our experimental comparison.

\myparagraph{ARF}
The Adaptive Range Filter (ARF)~\cite{alexiou2013adaptive} is based on a compactly-encoded binary tree whose leaves represent ranges of the universe and are associated with a flag indicating whether there is at least one key in that range.
Internal nodes allow navigating to the leaf containing the left endpoint $a$ of the query range $[a,b]$, and the leaves to its right are inspected until either one of them has a true flag, thus the answer is ``not empty'', or the leaf covering $b$ is reached and has a false flag, thus the answer is ``empty''.
ARF adapts to the data and query distribution by learning from false positive queries and adjusting its shape accordingly. 
As reported in \cite{zhangSurfPracticalRange2018}, ARF can be up to 1300$\times$ larger than \surf, described next, while also exhibiting a higher false positive rate. Thus, we do not further consider ARF.

\myparagraph{\surf}
The Succinct Range Filter (\surf)~\cite{zhangSurfPracticalRange2018,Zhang2020surfj,Zhang2021cacm} is built upon a com\-pact\-ly-encoded trie, called Fast Succinct Trie, that stores, for each key $s \in S$, the shortest prefix $p_s$ of $s$ such that $s$ can be uniquely identified among all the strings in $S$, followed by a number $m$ of \textit{suffix bits} following that key prefix.
For improving the filter performance on just point queries, these $m$ bits can also be set to a hash of the key.
A range emptiness query on $[a, b]$ is answered by looking in the truncated trie for the smallest key $k$ (which can include its suffix bits if the search reaches a leaf) such that $k$ is lexicographically $\ge a$. The result of the query is given by the result of the lexicographic comparison $k \le b$.
We use \surf in our experimental comparison.

\myparagraph{\rosetta}
The Robust Space-Time Optimized Range Filter (\rosetta) \cite{luoRosettaRobustSpacetime2020} consists of $\log u$ Bloom filters organised in levels. For every key, each of its prefixes of length $k=1,\dots,\log u$ is inserted into the Bloom filter at level $k$.
A range emptiness query is answered by decomposing the query range into dyadic intervals, which are used to probe the corresponding Bloom filters.
If all the probes return a negative answer, the query range is empty. Otherwise, each range that returned a positive answer is recursively decomposed at the next level.
In practice, \rosetta tunes itself to minimise the false positive rate under a given space budget by allocating more bits to the Bloom filters that are probed more frequently, based on a sample of the query workload.
We use Rosetta in our experimental~comparison.

\myparagraph{\snarf} The Sparse Numerical Array-Based Range Filter (SNARF) \cite{vaidyaSNARFLearningenhancedRange2022} uses a bit array $B$ of $K n$ bits, where $K$ is a suitably large parameter that impact on the false positive rate, and a function $f(x) = \lfloor \text{MCDF}{(x)} \cdot K n\rfloor$, where MCDF is a monotonic estimate of the CDF of the keys in $S$, i.e. $\lfloor \text{MCDF}(x) \cdot n \rfloor$ is an estimate of the rank of~$x$ in~$S$. The function $f$ is built by taking one key every $t$ keys in the sorted $S$, and using these key samples as endpoints of linear splines. The bit array $B$, initially empty, is filled with $B[f(x)]=\bone$ for each $x\in S$, and then compressed. A range emptiness query $[a, b]$ returns ``not empty'' if and only if there is at least a \bone-bit in the range $B[f(a), f(b)]$. 
We use \snarf in our experimental comparison.

\myparagraph{\proteus} \proteus~\cite{knorr2022proteus} combines the trie-based prefix filtering of the Fast Succinct Trie with the filtering of the Prefix Bloom Filter.
Differently from \surf, \proteus does not encode in the trie a unique prefix for every key but rather all unique key prefixes of a fixed length $l_1$, and it implements a single (prefix) Bloom filter for all key prefixes of length $l_2 > l_1$.
If the range emptiness query is not resolved after descending the trie up to the prefix length $l_1$, i.e. if there are matching leaves so that we cannot yet return ``not empty'', then the Prefix Bloom Filter is probed for each length-$l_2$ prefix extending the length-$l_1$ prefix of each matching leaf, returning ``empty'' if all these probes return false, ``not empty'' otherwise.
The values of $l_1$ and $l_2$ are determined by an algorithm that minimises the false positive rate given the input keys, a sample query workload, and a space budget.
We use \proteus in our experimental comparison.

\myparagraph{\bloomrf} The Bloom Range Filter (\bloomrf)~\cite{mossner2022bloomrf} hashes a key into a hash code composed of positions that are used to set bits to \bone in a bit array~$B$. 
The hash code is such that equal key prefixes have equal hash code prefixes (thus encoding range information in the hash code), and its position components preserve the order of prefixes (thus improving data locality).
A query range is decomposed into dyadic intervals whose emptiness is determined by checking in $B$ the appropriate bits computed via the hash code above.
We could not experiment with \bloomrf because its implementation is not yet open source,\footnote{Personal communication with the authors.} but we comment on it in the theoretical comparison of Section~\ref{sec:theoretical}.

\myparagraph{\rencoder} The Range Encoder (\rencoder)~\cite{Wang2023rencoder} too consists of a bit array~$B$, initially empty. It splits each input key into a 4-bit suffix~$s$ and the remaining prefix~$p$.
The suffix~$s$ is conceptually represented by a leaf in a complete binary tree with 16 leaves, whose nodes in the path from that leaf to the root are marked with a \bone, and the remaining nodes are marked with a \bzero.
Intuitively, nodes represent ranges of the universe and the bit marks record the presence of keys in a range.
The bit marks are then concatenated to form a 32-bit value, which is written into $B[h_i(p),h_i(p)+31]$ via an OR operation, where $h_i$ is a hash function, for $i=1,\dots,k$. The process is then repeated on the prefix $p$ of the key, and it stops when the whole key has been processed.
A query range is decomposed into dyadic intervals whose emptiness is determined via traversals of binary trees, which are recovered from $B$ via AND operations.
We use \rencoder in our experimental comparison.

\medbreak\noindent We conclude this section by mentioning that the problem of supporting efficient in-place insertions has only been touched upon in the literature. Indeed, current range filters are difficult to update efficiently due to their use of static compactly-encoded tries (\surf and \proteus), or learned functions and compressed bitvectors (\snarf).
Some other range filters like Prefix Bloom Filters, \rosetta, \bloomrf and \rencoder, instead, could be easier to update with new keys due to their design (loosely) based on Bloom filters, but the impact of insertions on the false positive rate has not yet been explored.  %
Since in this paper we do not deal with these issues, we leave it as an open problem~\cite{Dayan2023infinifilter}.

\section{Grafite: An Optimal Range Filter}
\label{sec:grafite}

We now introduce Grafite, which eventually solves the lack of robustness in state-of-the-art range filters. 
We start from the idea of Goswami~et~al.~\cite{goswamiApproximateRangeEmptiness2014} to solve the approximate range emptiness problem through hashing, and we take this idea into a simpler, practical and yet more succinct solution that is closer to the lower bound of Theorem~\ref{thm:lowerbound}.

\myparagraph{Hashing input keys}
Recall we are given a set $S$ of $n$ keys in a universe $[u]=\{0,\dots,u-1\}$, a false positive probability~$\varepsilon$, and an upper bound $L$ on the query range size.

Set $r = n L / \varepsilon$, and let $q \colon [u/r] \to [r]$ be a hash function taken from a \textit{pairwise-independent family} $Q$, i.e. a set of hash functions $Q = \{ q \colon  [u/r] \to [r]\}$ such that, for any pair of distinct keys $(x_1, x_2) \in [u/r]^2$ and any pair of (not necessarily distinct) hash codes $(y_1, y_2) \in [r]^2$, we have
$$\Pr_{q\in Q} \left[q(x_1) = y_1 \land q(x_2) = y_2\right] = \frac{1}{r^2}.$$
The simplest technique for constructing such a hash function~\cite{wegmanNewHashFunctions1981} is to select a large prime number $p > r$ and two random numbers $c_1, c_2 < p$ such that $c_1 \neq 0$, and then define the hash function as $q(x) = ((c_1 x + c_2) \bmod{p}) \bmod{r}$.

In addition to $q$, we define a hash function $h$ that preserves the {\em locality} of the hashed items and has a small collision probability~\cite{goswamiApproximateRangeEmptiness2014}:
\begin{equation}
    \label{eq:hash}
    h(x) = \left(q{\left(\lfloor x/r \rfloor\right)} + x \right) \bmod{r}.
\end{equation}

We use $h$ to transform the set $S=\{x_1, \dots, x_n\}$ of input keys from the original universe $[u]$ to the set $h(S)=\{h(x_1), \dots, h(x_n)\}$ of hash codes in the reduced universe $[r]$, and then we store $h(S)$ via a compact \textit{non-approximate} range emptiness data structure.

We will describe the data structure in a moment. For now, we notice that because of the hash function $h$, a range emptiness query $[a, b] \cap S \neq \emptyset\,?$ can be answered by verifying the existence of a value $h(x) \in h(S)$ such that
\begin{equation}
    \label{eq:conditions}
    \begin{cases}
    h(a) \leq h(x) \leq h(b) & \text{if } h(a) \leq h(b),\\
    h(x) \leq h(b) \lor h(x) \geq h(a)              & \text{otherwise.}
    \end{cases}
\end{equation}
The first case is straightforward, while the second one occurs if there is an \textit{overlap} of the hashed endpoints (i.e. $h(a) > h(b)$) as a consequence of the modulo and the reduced universe.
If a value $h(x)$ which satisfies~\eqref{eq:conditions} is found, we answer ``not empty''. Otherwise, we answer ``empty''.\footnote{In the special case the query range $[a, b]$ crosses one of the $u/r$ boundaries, i.e. $\lfloor a/r \rfloor$ = $\lfloor b/r \rfloor - 1$, we split it into $[a, a']$ and $[b', b]$, where $a'$ is the largest value such that $\lfloor a'/r \rfloor = \lfloor a/r \rfloor$ and $b'$ is the smallest value such that $\lfloor b'/r \rfloor = \lfloor b/r \rfloor$, namely, $b' = b - (b \bmod r)$ and $a'=b'-1$. Then, the two range emptiness queries are solved by applying the conditions~\eqref{eq:conditions}.}

It should be clear that there can be no false negatives. For the false positives, there is the following result (which is a straightforward generalisation of a result in \cite{goswamiApproximateRangeEmptiness2014}).

\begin{lemma}[\cite{goswamiApproximateRangeEmptiness2014}]
\label{th:lemma_are}
    \looseness=-1
    The approach based on the hash function~\eqref{eq:hash} and the conditions~\eqref{eq:conditions} guarantees a false positive probability of at most~$\varepsilon$ for query ranges of size $L$, and at most $\ell \varepsilon/L$ for ranges of size~$\ell \leq L$. 
\end{lemma}
\begin{proof}
A false positive occurs when no key in $S$ is in the query range $I$ but there is a hash collision between a key $x \in S$ and a point $y\in I$.
From~\cite[Lemma~3.1]{goswamiApproximateRangeEmptiness2014}, such a collision happens with probability ${\Pr[h(x) = h(y)]\leq1/r}$.
The false positive probability is then given by a union bound over all possible collisions between keys in $S$, which are $n$, and points in $I$, which are $\ell\leq L$, thus it is
\[
    \sum_{x \in S}{\sum_{y \in I}{\Pr\left[h(x) \!=\! h(y)\right]}}
    \leq \sum_{x \in S}{\sum_{y \in I} \frac{1}{r}} = \frac{n\ell}{r} = \frac{n\ell}{\tfrac{nL}{\varepsilon}} = \frac{\ell \varepsilon}{L} \leq \varepsilon.\qedhere
\]

\end{proof}

\myparagraph{Storing hash codes succinctly}
Having defined how approximate range emptiness can be achieved through hashing, the following step is to store the hash codes $h(S)$.
Goswami~et~al.~\cite{goswamiApproximateRangeEmptiness2014} store the  hash codes together with a sophisticated prefix search data structure from~\cite{Belazzougui2010weakprefix} to check hash codes for inclusion in a query range.
Our study proposes a much simpler data structure that builds on the classic Elias-Fano integer code~\cite{eliasEfficientStorageRetrieval1974,fanoNumberBitsRequired1971} together with an efficient procedure to check hash codes for inclusion in a range, explained below.
As we will show, we will obtain a practical range filter, which actually has an even better space than the solution of~\cite{goswamiApproximateRangeEmptiness2014}.

\begin{figure*}
    \centering
    \includegraphics[scale=0.97]{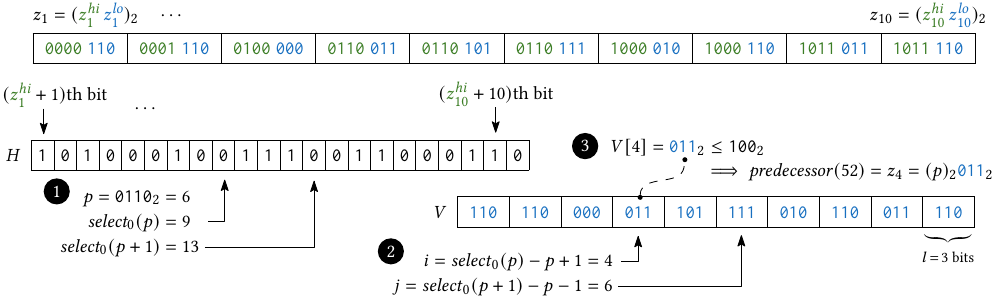}
    \caption{\boldmath  An example of Grafite storing the compressed hash codes $6, 14, 32, 51, 53, 55, 66, 70, 91, 94$ (see Example~\ref{ex:encoding}), and some steps needed for answering a range emptiness query (see Example~\ref{ex:query}).}
    \label{fig:example}
\end{figure*}

Let $z_1,\dots,z_n$ be the deduplicated sorted set of hash codes in $h(S)$.\footnote{Due to collisions there can be fewer than $n$ distinct hash codes, but we prefer using~$n$ in our description and bounds for simplicity. The space without the $k$ duplicates is $k \log\tfrac{L}{\varepsilon}$ bits lower, where the expected value of $k$ can be shown to be $\varepsilon(n-1)/(2L)$ with classic hash tables analyses under the simple uniform~hashing~assumption~\cite{clrs4thEd}.\label{foot:collisions}}
We split the length-$\lceil \log r \rceil$ binary representation of each $z_i$ into a low part $\lo{z_i}$ consisting of the $l = \lfloor \log\tfrac{r}{n} \rfloor = \lfloor \log\tfrac{L}{\varepsilon} \rfloor$ least significant bits of $z_i$, and a high part $\hi{z_i}$ consisting of the remaining $\lceil \log r \rceil-l$ most significant bits of $z_i$.
The low parts are concatenated into a vector $V[1,n]$ of $l$-bit cells, which thus takes $nl = n \lfloor \log\tfrac{L}{\varepsilon} \rfloor$ bits overall.
The high parts are encoded in a bitvector $H[1,\hi{z_n}+n+1]$ where the positions $\hi{z_i}+i$ are set to 1, and the remaining positions are set to 0. 
This completes the succinct encoding of $h(S)$, whose bit-size can be shown to be upper bounded by $n \log\tfrac{L}{\varepsilon} + 2n$.

\begin{example}\label{ex:encoding}
  Suppose we are given the set $S=\{9, 48, 50, 191,\allowbreak 226, 269, 335, 446, 487, 511\}$ with $n=10$ keys, and the values $L=4$, $\varepsilon=0.4$, thus giving $r=nL/\varepsilon=100$. Say we use the hash function $h(x) = (q(\lfloor x/r \rfloor) + x ) \bmod r$, where we choose $q(x)=((c_1 x + c_2) \bmod p) \bmod r$ with parameters $p=2^{31}-1$, $c_1=10$, and $c_2=5$, thus giving the set of hash codes $h(S)=\{14, 53, 55, 6, 51,\allowbreak 94, 70, 91, 32, 66\}$.  
  Figure~\ref{fig:example} shows the binary representation of the integers $z_1=6, \dots, 94=z_n$ corresponding to the sorted set $h(S)$ at the top (where the low $l=\lfloor \log\tfrac{L}{\varepsilon}\rfloor =3$ bits and the remaining $4$ bits are highlighted with different colours) and their Elias-Fano encoding via the vectors $H$ and $V$ at the bottom.
\end{example}

\myparagraph{Searching hash codes efficiently}
We now describe how to search within the hash codes $h(S)$ so that both conditions in~\eqref{eq:conditions} can be checked efficiently.

For the first branch in~\eqref{eq:conditions}, we need to augment the Elias-Fano encoding with an operation that  checks for the existence of an $h(x)$ such that $h(a) \leq h(x) \leq h(b)$.
To this end, we use the well-known $\mathit{predecessor}(y)$ operation, which given $y \in [r]$, returns the largest element $z_k$ smaller than or equal to $y$~\cite{Navarro:2020pred}. 
We first compute $z_k = \mathit{predecessor}(h(b))$ and then check if $z_k \geq h(a)$. If this is the case, then there is at least a hash code $z_k=h(x)$ in the range $[h(a), h(b)]$, where $x \in S$. Thus the first branch is satisfied, and the answer to the approximate range emptiness query is ``not empty''. If not, i.e. $z_k < h(a) \leq h(b)$, then the first branch is not satisfied and the answer is ``empty''.

\label{par:predecessor}
The $\mathit{predecessor}(y)$ operation is implemented by first identifying the range of hash codes $z_i,\dots,z_j$ that share the same high part $\hi{y}$ of $y$, and then by binary searching in the subarray $V[i,j]=[\lo{z_i}, \dots, \lo{z_j}]$ for the predecessor of $\lo{y}$.
Since $V[i,j]$ might contain at most $\cramped{2^l}$ configurations of $l$-bit integers, the binary search runs in $O(\log 2^l)=O(\log \tfrac{L}{\varepsilon})$~time.
Let us detail these steps for completeness~\cite{OkanoharaS07,Vigna2013quasi,ottaviano2014partitioned,Navarro:2016book}.
For identifying the range~$[i,j]$, we need the $\sel_b(k)$ operation, which returns the position of the $k$th $b$-bit in $H$, for $b \in \{\bzero,\bone\}$ (we define $\sel_b(0)=0$). This operation can be implemented in $O(1)$~time using $o(|H|)=o(n)$~bits~\cite{Jacobson:1989,clark1997compact}.
Then, by definition of $H$, the hash codes  $z_i,\dots,z_j$ with the same high part $p=\hi{y}$ form a contiguous sequence of \bone-bits in the subarray $H[p+i,p+j]$ followed by a \bzero in $H[p+j+1]$.
Thus, the subarray $H[p+i,p+j+1]=\bone^{j-i+1}\bzero$ corresponds to the subarray $H[\sel_\bzero(p)+1,\sel_\bzero(p+1)]$, and hence the range $[i,j]$ is given by setting the endpoints of these subarrays equal, thus yielding $i=\sel_\bzero(p)-p+1$ and $j=\sel_\bzero(p+1)-p-1$.
Finally, in the corner case that the binary search in $V[i,j]$ finds that the predecessor of $y$ is $z_{i-1}$, we recover this element via a random access operation: the high part of $z_{i-1}$ is retrieved as $\hi{z_{i-1}}=\sel_1(i-1)-(i-1)$, and the low part is readily available as $\lo{z_{i-1}}=V[i-1]$.

For the second branch in~\eqref{eq:conditions}, it is enough to random access the smallest and largest value in $h(S)$, i.e. $z_1$ and $z_n$, and do the comparison $z_1 \leq h(b) \lor z_n \geq h(a)$.

Algorithms \ref{alg:constr} and \ref{alg:query} contain the pseudocode for the construction and query algorithms on Grafite, respectively. For the construction, we notice that $\textsc{BuildEliasFano}$ runs in linear time, while $\textsc{Sort}$ takes the time to sort $n$ integers of length $\lceil \log r \rceil$, for which there exist very efficient sequential and parallel algorithms~\cite{Axtmann2022sorting}.

\begin{example}\label{ex:query}
  Let us solve the range emptiness query for the range $[44,47]$ on the Grafite instance of Example~\ref{ex:encoding} and Figure~\ref{fig:example}. First, we compute $h(a)=h(44)=49$ and $h(b)=h(47)=52$. Then, since $h(a) \leq h(b)$, we need to compute $z_k=\mathit{predecessor}(h(b))$ and do the check $z_k \geq h(a)$. As depicted in Figure~\ref{fig:example}, we find such a $z_k$ by: {\small\circled{1}} taking the high part $p=\mathtt{0110}_2=6$ of $h(b)=52$ and computing $\sel_0(p)=9$ and $\sel_0(p+1)=13$; {\small\circled{2}} computing $i=\sel_0(p)-p+1=4$ and $j=\sel_0(p+1)-p-1=6$; {\small\circled{3}} doing a binary search in $V[i,j]$ for the predecessor of the low part $\mathtt{100}_2=4$ of $h(b)$, which yields $V[4]=\mathtt{011}_2$. This reveals that $\mathit{predecessor}(h(b))=z_4=51$, and since $z_4 \geq h(a) = 49$, we return ``not empty'', which is a false positive error since $[44,47] \cap S = \emptyset$.
\end{example}

Combining the above description of Grafite with Lemma~\ref{th:lemma_are}, and recalling the restriction on the range size $L \leq u\varepsilon/n$ (see  Theorem~\ref{thm:lowerbound} and the discussion below it), we can state the following result.

\begin{theorem}\label{thm:grafite}
Given a set of $n$ keys from a universe $[u]$, $\varepsilon \in (0,1)$, and $L \in [1,u\varepsilon/n]$, Grafite takes $n\log\tfrac{L}{\varepsilon}+2n+o(n)$ bits of space and answers approximate range emptiness queries in time $O(\log\tfrac{L}{\varepsilon})$ with a false positive probability of at most $\varepsilon$ for query ranges of size~$L$, and at most $\ell \varepsilon/L$ for query ranges of size~$\ell \leq L$.
\end{theorem}

Since $n\log\tfrac{L}{\varepsilon}+2n+o(n) = n\log\tfrac{L}{\varepsilon} + O(n)$, Grafite matches the space lower bound of the approximate range emptiness problem (Theorem~\ref{thm:lowerbound}), thus it is space-optimal, while offering constant time queries whenever $L/\varepsilon=O(1)$.

\SetFuncSty{textsc}
\SetKwBlock{Begin}{function}{end}
\SetKwProg{Fn}{function}{}{}
\SetInd{0.5ex}{1.8ex}
\DontPrintSemicolon
\setlength{\algomargin}{1.5ex}
\begin{algorithm}[t]
\SetKwFunction{FMain}{Construct}
\caption{Construction of Grafite.}\label{alg:constr}
\Fn{\FMain{$S[1,n]$, $L$, $\varepsilon$}}{
$G.r \gets nL/\varepsilon$\;
$G.h \gets \,$ The hash function (\ref{eq:hash})\;%
$V \gets \text{An empty array of size } n$\;
\For{$i = 1$ \text{{\bf to}} $n$}{
    $V[i] \gets G.h{(S[i])}$\;
}
$\textsc{Sort}{(V)}$\;
$G.\mathit{ef} \gets \textsc{BuildEliasFano}(V)$\;
\Return $G$\;
}
\end{algorithm}
\begin{algorithm}[t]
\SetKwFunction{FMain}{RangeEmptinessQuery}
\caption{Range emptiness query in Grafite.}\label{alg:query}
\Fn{\FMain{$G$, $a$, $b$}}{
$h_{a} \gets G.h{(a)}$\;
$h_{b} \gets G.h{(b)}$\;
\If{$h_{a} > h_{b}$}{
    \Return ``Not empty'' \textbf{iff} $G.\mathit{ef}.\mathit{first} \le h_{b} \lor G.\mathit{ef}.\mathit{last} \ge h_{a}$\;
}
\Return ``Not empty'' \textbf{iff} $G.\mathit{ef}.\mathit{predecessor}{(h_{b})} \geq h_{a}$ \;}
\end{algorithm}

Notice that Grafite has several important features.
First, the false positive probability is bounded \emph{regardless} of the input set $S$ and query workload, thus solving the first challenge faced by known practical range filters mentioned in Section~\ref{sec:intro}.
Second, the query time is independent of $n$ (and $u$), thus making Grafite efficient even for large sets of input keys.
Third, Grafite does not require any sophisticated tuning procedure, but it can be used out of the box by just specifying $\varepsilon$~and~$L$.

We stress that, after $L$ has been set, Grafite can answer on both query ranges of size $\ell$ smaller than $L$ (with a smaller chance of false positives than $\varepsilon$) and larger than $L$ (with a higher chance of false positives than $\varepsilon$), because the presence of $L$ in Theorem~\ref{thm:grafite} is technical and  serves to make the false positive probability $\leq \varepsilon$.
As a matter of fact, since the space usage is $\log\tfrac{L}{\varepsilon}+2$~bits per key,\footnote{The $o(1)$ term we omit here can be just $0.035$ bits per key in practice~\cite{kurpicz2022pasta}.} we can build Grafite by just setting the space budget to a constant~$\budget$, hence $\varepsilon = L/2^{\budget-2}$, and we can answer range emptiness queries with a false positive probability of at most
\begin{equation*}
      \frac{\ell\varepsilon}{L}
    = \frac{\ell L/2^{\budget-2}}{L}
    = \frac{\ell}{2^{\budget-2}},
\end{equation*}
in time
\begin{equation*}
      O\left(\log\frac{L}{\varepsilon}\right)
    = O\left(\log\frac{L}{L/2^{\budget-2}}\right)
    = O\left(\log 2^{\budget-2}\right)
    = O(\budget),
\end{equation*}
thus proving the following result (which solves the second challenge faced by known practical range filters mentioned in Section~\ref{sec:intro}).

\begin{corollary}\label{cor:grafite}
Given a set of $n$ keys and a budget of $\budget=O(1)$ bits per key, Grafite answers approximate range emptiness queries in $O(1)$~time with a false positive probability of at most $\min\{1,\ell/2^{\budget-2}\}$, where $\ell$ is the query range size.
\end{corollary}

Observe that, a similar derivation of Corollary~\ref{cor:grafite} with the data structure of Goswami~et~al.~\cite{goswamiApproximateRangeEmptiness2014} would lead to a false positive probability higher (actually, strictly higher, due to the lower-order terms we omit) than $\ell/ 2^{\budget-3}$, which is worse than the one achieved by Grafite (see also Section~\ref{sec:theoretical}).

Finally, we mention that instead of returning a boolean answer, Grafite can return an approximate count of the keys that intersect the given query range without any change in its space or query time complexity, thus potentially being a practical and efficient solution for this interesting problem too~\cite{Alstrup2001reporting}. It suffices to return the difference between the \emph{ranks} at the hashed endpoints of the query range (possibly adjusting the result with the expected number of collisions in the range, as per Footnote~\ref{foot:collisions}), where the rank of a hashed element can be found easily during the  \textit{predecessor} operation on the Elias-Fano sequence.

\section{Bucketing: A Heuristic Range Filter}
\label{sec:bucketing}

\looseness=-1
We now introduce a very simple heuristic range filter named Bucketing. Bucketing has the same weakness of known heuristic range filters, namely, it provides little or no filtering on correlated query workloads, thus its purpose is not to compete with Grafite, which instead provides robust and consistent filtering effectiveness regardless of the input set and query workload.
Rather, Bucketing will serve us to show experimentally that, on certain inputs experimented in the literature,  one does not need to resort to the sophisticated heuristic filter designs proposed in the literature, because simpler solutions can experimentally match or improve their filtering effectiveness while being more efficient to query and construct.

Given a set $S=\{x_1,\dots,x_n\}$ of $n$ keys in the universe~$[u]$, and given an integer $s \geq 1$, we split the universe into $u/s$ buckets of size $s$.
Then, we create a bitvector~$\bv$ of size $u/s$ that indicates with $\bone$ in position $i$ if there exists at least a key $x \in S$ that falls in the $i$th bucket.
That is, $\bv$ is initially empty, and we set $\bv[x/s] = \bone$ for each $x \in S$ (we omit floors for simplicity).

Let $t$ be the number of $\bone$-bits in $\bv$, which depends on the distribution of the input data.
Clearly, $t$ cannot be more than the size of~$\bv$ or than the number of elements in $S$, thus $t \leq \min\{u/s, n\}$.
By compressing $\bv$ with the Elias-Fano encoding, the total space occupancy is $t(\log\tfrac{u}{ts}+2)$~bits.
The construction can actually be done without creating $\bv$ by just considering the deduplicated list of the \bone-bit positions $x_1/s, \dots, x_n/s$. 

The parameter $s$ allows us to trade the space with the coarseness of such a lossy encoding of $S$.
Indeed, when ${s=1}$, we are losslessly encoding the input set (i.e. $t = n$) and the space is $n(\log\tfrac{u}{n}+2)$~bits, whereas if $s=u$ then a single bucket exists for the whole set (i.e. $t=1$) and its single entry in $\bv$ is \bone, thus the space is~0.

Similarly to Grafite (Section~\ref{par:predecessor}), we  augment the Elias-Fano encoding of $\bv$ with select data structures that occupy $o(t)$~bits and allow us to compute the $\mathit{predecessor}$ operation in $O(\log\tfrac{u}{ts})$~time. Then, given a query range~$[a, b]$, if  $\mathit{predecessor}(b/s) \geq a/s$ is true, then  $\bv[a/s, b/s]$ contains at least a $\bone$-bit and we answer ``not empty''. Otherwise, we answer ``empty''. 

It goes without saying that false negatives are not possible and that a false positive happens when $[a,b] \cap S = \emptyset$ but there is a key $k\in S$ such that $k<a$ and $k$ falls into bucket number $a/s$, or symmetrically if $k > b$ and $k$ falls into bucket number $b/s$.
Similarly to other heuristic filters, a bound on the false positive rate that holds regardless of the input data and query distributions cannot be proved.
Moreover, we expect this approach to provide no filtering as the correlation increases, due to endpoints of the query range falling in non-empty buckets. 

\section{Theoretical Comparison}
\label{sec:theoretical}
\begin{table*}[t]
\caption{\boldmath Summary of known and new theoretical results achieved by range filters. Recall that $n$ is the number of input keys from a universe of size $u$, $L$ is an upper bound on the query range size, and $\varepsilon$ is the false positive probability.}
\label{tab:theoretical}
\centering
\renewcommand{\arraystretch}{1.18}
\setlength{\tabcolsep}{12pt}
\begin{tabular}{l l l l c}
\cmidrule[\heavyrulewidth](l){2-5}
 &
   &
  \textbf{Space in bits} &
  \textbf{Query time} &
  \textbf{Practical impl.} \\ \cmidrule(l){2-5} 
 
 \ldelim\{{5}{*}[{\rotatebox[origin=c]{90}{Heuristic}}\;]\hspace{-1.8em}
 &
  \surf{}~\cite{zhangSurfPracticalRange2018} &
  $(10+m)n + 10z + o(n+z)$* &
  $O(\log u)$ &
  $\checkmark$ \\
 &
  \snarf{}~\cite{vaidyaSNARFLearningenhancedRange2022}  &
  $n\log K  + 2.4n$\textsuperscript{\textdagger} &
  $\Omega(\log{n})$ &
  $\checkmark$ \\
 &
  \proteus{}~\cite{knorr2022proteus} &
  ? &
  ? &
  $\checkmark$ \\
 &
  \bloomrf{}~\cite{mossner2022bloomrf} &
  ? &
  $O(\log\tfrac{u}{n})$ &
  $\checkmark$ \\ 
 &
  Bucketing (this paper) &
  $t\log\tfrac{u}{ts}+2t + o(t)$\textsuperscript{\textdaggerdbl} & 
  $O(\log\tfrac{u}{ts})$ & 
  $\checkmark$ \\ \cmidrule(l){2-5} 
 \ldelim\{{4}{*}[{\rotatebox[origin=c]{90}{FPR-bounded}}\;]\hspace{-1.8em}
 &
  Theoretical baseline (cf. Section~\ref{sec:related}) &
  $n \log \tfrac{L}{\varepsilon} + O(n)$ &
  $O(L)$ &
   \\
 &
  Goswami et al.~\cite{goswamiApproximateRangeEmptiness2014} &
  $n \log\tfrac{L}{\varepsilon} + 3n + o(n\log\tfrac{L}{\varepsilon})$ &
  $O((\log\tfrac{nL}{\varepsilon})/w)$ &
   \\
 &
  \rosetta{}~\cite{luoRosettaRobustSpacetime2020} &
  $1.44 \cdot n \log\tfrac{L}{\varepsilon}$ &
  $\Omega((\log L) \log(2-\varepsilon))$\textsuperscript{\S} &
  $\checkmark$ \\
 &
  Grafite (this paper) &
  $n\log\tfrac{L}{\varepsilon}+2n+o(n)$ &
  $O(\log \tfrac{L}{\varepsilon})$ &
  $\checkmark$ \\ \cmidrule(l){2-5}
 & 
  \textbf{Lower bound} (Theorem~\ref{thm:lowerbound}) &
  $n \log{\left(\frac{L^{1-O(\varepsilon)}}{\varepsilon}\right)} - O(n)$ &
   &
   \\ \cmidrule[\heavyrulewidth](l){2-5} 
\end{tabular}

\begin{minipage}{0.85\textwidth}
\footnotesize
\begin{itemize}
\item[*] \!Using the space-efficient LOUDS-Sparse encoding, where $z$ is the number of internal nodes, and $m$ is the given number of suffix bits.
\item[\textsuperscript\textdagger] \!$K\geq 1$ is a suitably large parameter of SNARF.
\item[\textsuperscript\textdaggerdbl] \!$s$ is a positive integer parameter (higher values encode the input set more coarsely), and $t \in [1,\min\{n,u/s\}]$ depends on the input data.
\item[\textsuperscript\S] \!Expected time. Worst-case time is $O(L\log\tfrac{1}{\varepsilon})$.
\end{itemize}
\end{minipage}
\end{table*}

We now compare the space-time bounds of Grafite (Theorem~\ref{thm:grafite}) with those of the state-of-the-art range filters discussed in Section~\ref{sec:related}.
We distinguish two kinds of range filters, the ones that provide a \emph{bounded} false positive probability~$\varepsilon$ thus solving the approximate range emptiness problem formulated in Section~\ref{sec:intro}, and the \emph{heuristic} ones, which do not provide any guarantee unless some  assumptions on the input data and query distribution are met.
Therefore, the space complexity of these latter range filters cannot and will not be compared with that of Grafite (unless under said assumptions).

Table~\ref{tab:theoretical} summarises known and new bounds.
Some complex time bounds are simplified with the $\Omega$-notation, which still allows comparing them with Grafite.
All time bounds do not include the $O((\log u)/w)$ time to process the two endpoints of the query range, which is typically neglected because any solution has to read them.

\myparagraph{Goswami~et~al.'s solution}
The data structure proposed for the $w$-bit word RAM model by Goswami~et~al. takes $n \log\tfrac{L}{\varepsilon} + O(n) + o(n\log\tfrac{L}{\varepsilon})$ bits, where the $O(n)$ term hides $3n+o(n)$~bits~\cite[\S3.2]{goswamiApproximateRangeEmptiness2014}.
Hence, Grafite is better in space than this data structure by an additive term $n + o(n\log\tfrac{L}{\varepsilon})$, thus it is closer to the space lower bound of approximate range emptiness data structures (Theorem~\ref{thm:lowerbound}).

On the other hand, the query time of Grafite is $O(\log\tfrac{L}{\varepsilon})$ while the query time of Goswami~et~al.'s data structure is $O((\log\tfrac{nL}{\varepsilon})/w)$.
The former is higher than the latter when $n = O((L/\varepsilon)^{w-1})$.
In the case $L/\varepsilon=O(1)$, then queries in both data structures take $O(1)$ time because it is usually assumed $w=\Omega(\log n)$.

\myparagraph{\rosetta} Rosetta allows tuning the false positive probability of its per-level Bloom filters.
We use the tuning from~\cite[\S3.1]{luoRosettaRobustSpacetime2020} that achieves approximately $1.44 \cdot n \log\tfrac{L}{\varepsilon}$ bits of space by setting the probability of false positives to $\varepsilon$ for the last-level Bloom filter and to $1/(2-\varepsilon)$ for each other upper-level Bloom filter.
The space of Grafite is better, since $1.44 \cdot \log\tfrac{L}{\varepsilon} < \log\tfrac{L}{\varepsilon} + 2$ if and only if $L<23.36\varepsilon$.

\looseness=-1
For the query time, the worst-case number of Bloom filter probes done by Rosetta is $O(L)$ and the expected number is $\Omega(\log L)$, as per the analysis in~\cite[\S3.2]{luoRosettaRobustSpacetime2020}.
The probe time of the last-level Bloom filter is $\Theta(\log\tfrac{1}{\varepsilon})$, which is higher than the $\Theta(\log(2-\varepsilon))$ probe time of each other upper-level Bloom filter.
So the worst-case query time of Rosetta is $O(L\log\tfrac{1}{\varepsilon})$, which is worse than the query time of Grafite, and the expected query time of Rosetta is $\Omega((\log L) \log(2-\varepsilon))$, which is equivalent to the query time of Grafite if $\varepsilon$ is a constant.

\myparagraph{\surf}
Let $z$ be the number of internal nodes in the Fast Succinct Trie at the core of \surf, and recall it stores one leaf and $m$ suffix bits for each of the $n$ input keys.
The trie uses the LOUDS-Dense encoding for the upper levels and LOUDS-Sparse for the lower levels.
Following~\cite[\S2.5]{zhangSurfPracticalRange2018}, we assume the more space-efficient LOUDS-Sparse encoding is used, in which each node takes 10 bits.
Considering the $o(z+n)$ bits for the rank/select data structures, the total space sums up to $n m + 10(n+z) + o(n+z) = (10+m)n + 10z + o(n+z)$ bits. 
From this analysis (confirmed by experiments), we  infer that \surf needs at least 10 bits per key, which can be restrictive in applications with a low space budget.

The query time of \surf is given by the time to traverse the trie and then compare the suffix bits. Thus, for a trie of height $h$, the time is $O(h)$ if the suffix bits fit into a machine word (thus they can be accessed in constant time), and if each branching step takes constant time, e.g. because the trie has a constant bounded fan-out.
Since the input keys are of length $O(\log u)$, the query time is $O(h)=O(\log u)$.
Even for a fairly large $L \leq u\varepsilon/n$ (cf. Theorem~\ref{thm:grafite}), Grafite is faster than \surf because it has query time $O(\log\tfrac{L}{\varepsilon})=O(\log\tfrac{u}{n})$.

\myparagraph{\snarf}
\snarf was shown to take approximately $n\log K  + 2.4n$ bits, where $K$ is a suitably large parameter impacting on the false positive rate \cite[\S5]{vaidyaSNARFLearningenhancedRange2022}.
For the query time, the paper does not give a precise analysis, but we notice that it requires performing a binary search on the sample of $n/t$ keys (where $t$ is a constant) to identify the correct spline model, followed by decoding the compressed bit array. Due to the binary search, \snarf takes time $\Omega(\log \tfrac{n}{t})=\Omega(\log n)$, which is already at least asymptotically the query time of Grafite.
Indeed, Grafite binary searches on a range with $\min\{n, L/\varepsilon\}$ keys. 

Under the assumption of uniform keys, uniform query workload, and $u \gg nK$, \snarf was shown to have a false positive probability of $(u/K)/(u-nL)$ for query ranges of size $L$~\cite[\S3]{vaidyaSNARFLearningenhancedRange2022}.
This is approximated to $1/K$ under the additional assumption of $nL \ll u$.
In such a restricted setting, \snarf takes $\log{L}-0.4$ bits per key less than Grafite with $\varepsilon$ set to $1/K$.
On the flip side, SNARF suffers a high false positive rate in correlated workloads (see Section~\ref{ssec:exp-robustness}).

\myparagraph{\proteus} 
The \proteus paper~\cite{knorr2022proteus} does not provide a closed formula for the space and the query time taken by this data structure.
Indeed, \proteus tunes its configuration parameters $(l_1,l_2)$ via an algorithm whose inputs are the keys, a query workload, and a space budget (cf.~\cite[Alg. 1]{knorr2022proteus}). This makes it difficult to provide a satisfactory space bound other than for the extreme configurations that turn it into either a full Fast Succinct Trie, or a full Prefix Bloom Filter.

For the query time, we could not derive a satisfactory analysis either, but we observe that \proteus uses a Fast Succinct Trie on prefixes of uniform depth $l_1$ and a Prefix Bloom Filter for prefixes of length $l_2 > l_1$, thus it might require a trie traversal plus several queries to the Prefix Bloom Filter (our experiments will show that \proteus is much slower than Grafite).

\myparagraph{\bloomrf}
The authors of \bloomrf build a model of the false positive rate given a space budget in \cite[\S5--6]{mossner2022bloomrf}.
We prefer to not report this model here due to its complicated design and assumptions, but we content ourselves to notice that it is influenced by the input data distribution (cf. the constant~$C$ in \cite{mossner2022bloomrf}), thus making \bloomrf a heuristic solution.

The query time of \bloomrf is given by the time to compute $k=\lceil \log\tfrac{u}{n} / \Delta \rceil$ hash functions, where $\Delta\geq 1$ is a parameter of the data structure~\cite[\S6]{mossner2022bloomrf}. Thus \bloomrf has query time $O(\log\tfrac{u}{n})$, which is no better than  Grafite for the same considerations we make above for \surf.

\myparagraph{\rencoder} The authors of \rencoder show in~\cite[\S4]{Wang2023rencoder} that, under some assumptions, a false positive probability of $\varepsilon$ can be obtained using $O(n(k+\log\tfrac{1}{\varepsilon}))$~bits of space, where $k$ is the number of hash functions used in \rencoder.
This result is hard to compare with Grafite due to the lack of $L$ in the space bound (which seems to conflict with the lower bound of Theorem~\ref{thm:lowerbound}), due to the big-$O$, and due to the use of $k$, which also impacts on the query time (no precise indications on how to set $k$ are given).
In any case, the analysis in \cite[\S4.C]{Wang2023rencoder} suggests that \rencoder too is affected by correlated workloads, which is confirmed by our experiments of Section~\ref{ssec:exp-robustness}.

\section{Experiments}\label{sec:experiments}

We now perform the largest experimental comparison among range filters, both in terms of dataset size and in the number of tested solutions, and we show that:
\begin{enumerate}[leftmargin=*]
  \item The vast majority of existing range filters provide no filtering or much degraded filtering and query performance in the case of correlated query workloads. Instead, Grafite is among the (few) robust range filters, and it offers the overall best false positive rate (FPR) and query time already starting from mildly correlated query workloads.

  \item On uncorrelated query workloads, Bucketing offers, simultaneously, a filtering effectiveness that is very close to or better than the one achieved by the best-performing heuristic range filters, 5--13$\times$ faster queries, and 5--24$\times$ faster construction than them.
  
  \item Among robust range filters, Grafite is the best choice because it offers, simultaneously, the best FPR by up to 5 orders of magnitude, 9--92$\times$ faster queries, and 4--10$\times$ faster construction.
\end{enumerate}

Then, we conclude this experimental section by summarising our findings in terms of recommendations on which range filter to adopt for an application (Section~\ref{ssec:recommendations}).

\subsection{Experimental Setup}
\label{subsec:setup}

All the experiments are run on a machine equipped with a 1.80 GHz Intel Xeon E5-2650Lv3 CPU and 64 GB of RAM. The code of Grafite and the competitors is in C++ and is compiled with gcc-11. 
Our source code is available at \url{https://github.com/marcocosta97/grafite}.

\myparagraph{Competitors}
As motivated in Section~\ref{sec:related}, we compare Grafite and Bucketing with the following state-of-the-art range filters: \surf~\cite{zhangSurfPracticalRange2018},  \rosetta~\cite{luoRosettaRobustSpacetime2020}, \snarf~\cite{vaidyaSNARFLearningenhancedRange2022},\footnote{During the experiments we found that \snarf returns some false negatives, which is contrary to the definition of (range) filter. The authors believe this is due to computation overflows in the learned model, recommending the use of \snarf on significantly smaller input numbers (personal communication, Kapil Vaidya, 17 April 2023). We still experiment with \snarf, with the hope the false negatives do not affect its performance.\label{foot:snarf_bug}} \proteus~\cite{knorr2022proteus} and \rencoder~\cite{Wang2023rencoder}, including its variants \rencoderse and \rencoderss. This makes our study the largest one in terms of number of considered competitors.

\rosetta, \proteus and \rencoderse are auto-tuned on a sample of the queries with the procedures designed by the respective authors.
For \surf, we use real suffixes when testing against range queries and hashed key suffixes when testing against point queries, as suggested by \cite{zhangSurfPracticalRange2018}. 

\myparagraph{Datasets} 
We use synthetic and real-world datasets used in previous range filters evaluations \cite{zhangSurfPracticalRange2018, luoRosettaRobustSpacetime2020, vaidyaSNARFLearningenhancedRange2022,knorr2022proteus,Wang2023rencoder}:
\begin{itemize}[leftmargin=*]
    \item \textsc{Uniform}: 200M keys chosen uniformly at random from $[0, 2^{64})$.
    \item \textsc{Books}: 200M keys representing Amazon book sale popularity.
    \item \textsc{Osm}: 200M coordinates of locations from Open Street Map.
\end{itemize}
 
By using up to the entire dataset to build the range filters, we double the scale of the previously largest evaluation~\cite{vaidyaSNARFLearningenhancedRange2022}.

We build each range filter with space budgets ranging between $\approx$\,8 and 28 bits per key, which covers a large spectrum of trade-offs~\cite{luoRosettaRobustSpacetime2020,Wang2023rencoder}.\footnote{\surf takes no less than 10 bits per key, as per our analysis in Section~\ref{sec:theoretical}, and some configurations are not shown due to crashes. Some configurations of \proteus results in the same design thus giving overlapped points in our figures.}
We ensure that the space of a range filter does not exceed an explicit encoding of the input keys, namely $\log\tfrac{u}{n}+2$~bits per key via an Elias-Fano encoding, since this approach would solve the problem without false positives as discussed after Theorem~\ref{thm:lowerbound}.

\begin{figure*}
    \centering
    \includegraphics[width=\linewidth]{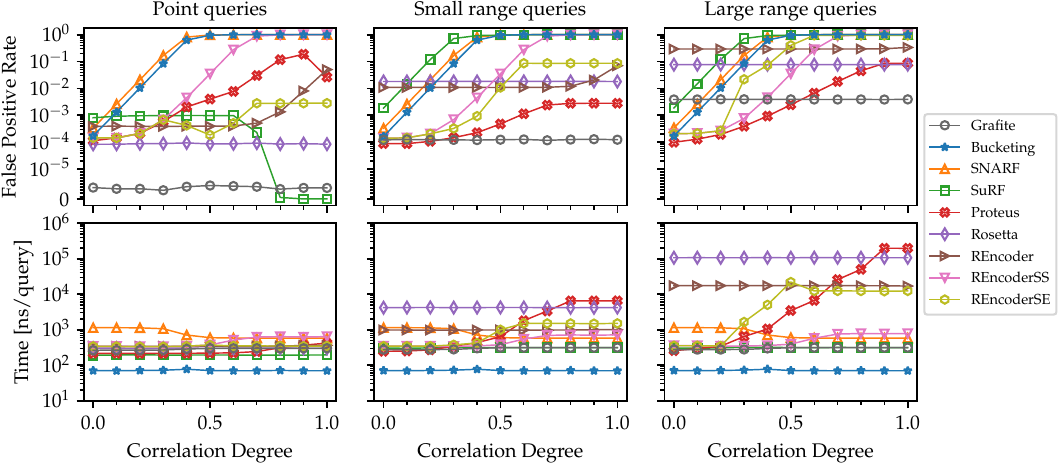}
    \caption{The majority of range filters provide no filtering (Bucketing, \snarf, \surf, \rencoderss) or much degraded filtering and query performance (\proteus, \rencoderse) as the key-query correlation increases. An adversary could exploit this weakness to make an attack on the availability of the system employing these heuristic range filters. Instead, Grafite and \rosetta are robust range filters, while \rencoder is robust for large range queries. Grafite offers significantly better query time and FPR than \rosetta and \rencoder.}
    \label{fig:corr_test}
\end{figure*}

\myparagraph{Query workloads}
Following the literature~\cite{vaidyaSNARFLearningenhancedRange2022,knorr2022proteus,luoRosettaRobustSpacetime2020,Wang2023rencoder}, we execute 10M range emptiness queries of the form $[x,x+L-1]$ in a single thread.
We distinguish between batches of \emph{point queries} in which $L=2^0$, \emph{small range queries} in which  $L=2^{5}$, and \textit{large range queries} in which $L=2^{10}$.

For the \emph{synthetic} dataset (\textsc{Uniform}), the left endpoint $x$ is chosen according to the following strategies: 
\begin{itemize}[leftmargin=*]
    \item \textsc{Uncorrelated}: $x$ is chosen uniformly at random from $[0, 2^{64})$.

    \item \textsc{Correlated}: a key $k$ is chosen uniformly at random from the dataset, and then $x$ is chosen uniformly at random from $[k, k + 2^{30 (1 - D)}]$, where $D$ is the \emph{correlation degree} that ranges from 0 (uncorrelated) to 1 (correlated)~\cite{vaidyaSNARFLearningenhancedRange2022}. If not explicitly varied, we set $D=0.8$.
\end{itemize}

For the \textit{real} datasets, the left endpoint $x$ is a key extracted (and removed) from the dataset. Notice that this query workload may be a mix of correlated and uncorrelated query ranges, depending on the distribution of the original input keys, from which $x$ is extracted.

In all the above strategies, we enforce the generation of \emph{empty} queries by discarding the query ranges that intersect the dataset.
This way, we evaluate the false positive rate (FPR) as the ratio between the number of ``not empty'' answers and the size of the batch.
In a separate experiment, we also test the query time of range filters on \emph{non-empty} queries. Note that the query time does not include the time to access a slow resource, such as a disk or a network drive, where the dataset might be stored. This time can vary greatly depending on the FPR and the hardware, or it might even be absent if the application requires no further check of a ``not empty'' answer (i.e. checking whether it is a true positive or not).

\myparagraph{Other datasets and query workloads}
We ought to report that we have also experimented with (i)~a dataset generated from a normal distribution (mean of $2^{63}$, standard deviation of $2^{64} \times 0.1$, which allows covering the universe and generating large range queries) in combination with the  \textsc{Uncorrelated} and \textsc{Correlated} query workloads, and (ii) the~\textsc{Uniform} dataset in combination with a normal query workload.
In all these cases, consistently with previously published evaluations~\cite{knorr2022proteus,vaidyaSNARFLearningenhancedRange2022}, we found no interesting change in the relative performance of range filters compared to using \textsc{Uniform} only, so we do not show them.

We have also experimented with the \textsc{Fb} dataset used in~\cite{Wang2023rencoder,vaidyaSNARFLearningenhancedRange2022,knorr2022proteus} but we found it to be too simple to be included in our evaluation because the mean value of the keys is $\approx 2^{38}$, and if we exclude the last 21 keys (that are larger than $2^{38}$), then an Elias-Fano encoding of the dataset would provide no false positives in just $\log\tfrac{2^{38}}{200 \cdot 10^6} + 2 \approx 12$ bits per key.
Indeed, we report that Grafite, due to its optimal design, provides an FPR of 0 on \textsc{Fb} when given a budget of only 12 bits per key, while the other range filters may still give false positives (as shown also in the papers above).

\subsection{Robustness of Range Filters}\label{ssec:exp-robustness}

Our first experiment aims to differentiate robust range filters from heuristic ones, thereby emphasising the necessity of treating them separately due to their distinct guarantees.
We consider the \textsc{Uniform} dataset and the \textsc{Correlated} query workload where the correlation degree $D$ is varied from 0 to 1, using a space budget for the range filters fixed to 20 bits per key.

The results in Figure~\ref{fig:corr_test} show that the FPR of Grafite and \rosetta is not affected by correlation, so we classify them as robust range filters. Grafite offers a better FPR than \rosetta by up to two orders of magnitude.
The FPR of \rencoder is affected by correlation, but this effect diminishes for larger range sizes. Grafite offers a better FPR than \rencoder by up to four orders of magnitude.

The FPR of \proteus too suffers from increased correlation, but it does not reach 1. In the case of slightly correlated (i.e. $D<0.5$) large range queries, \proteus shows a smaller FPR than Grafite, while Grafite has a better FPR in all the other cases.
We stress that \proteus is auto-tuned on the input keys and the query workload, so it has an advantage due to overfitting. In applications where the workload shifts, it might not retain this advantage. 

The FPR of \surf, \snarf and Bucketing approaches 1 for correlation degrees beyond 0.4, thus failing to provide any kind of filtering (the drop of FPR of \surf in point queries is expected because it ends up comparing hashed key suffixes).
The same holds for \rencoderss and \rencoderse for correlation degrees beyond 0.7 (the latter in the case of large range queries).

For what concerns the query time, Grafite is the fastest effective range filter across the various query range sizes and correlation degrees (Bucketing is the fastest range filter, but it is not always effective, as commented above).
The query time of \proteus, \rosetta and \rencoder increases for increasingly large query ranges, up to about 3 orders of magnitude more with respect to Grafite.
The query time of \proteus, \rencoderse and \rencoderss is affected by the correlation degree, which is another reason to classify them as non-robust.

In summary, with the exception of Grafite, \rosetta and, to a lesser extent, \rencoder, the vast majority of range filters provide \emph{no} filtering (\snarf, \surf, \rencoderss) or \emph{much degraded} filtering and query performance (\proteus, \rencoderse) in the case of correlated query workloads.
This is a significant concern given the importance of these workloads in applications that care about the local properties of data~\cite{luoRosettaRobustSpacetime2020} or given that malicious users could exploit this weakness to increase the network or disk accesses the range filters are deployed to prevent, thus posing a \emph{risk} on the availability of a data system.
Instead, Grafite is the overall best range filter in terms of FPR and query time already starting from mildly correlated query workloads, independently of the query range size.

Given the large number of competitors and the widely different guarantees they provide, our next experiments will focus separately on \emph{heuristic range filters}, namely \snarf, \surf, \proteus, \rencoderss, and \rencoderse, and on \emph{robust range filters}, namely Grafite, \rosetta, and \rencoder.

\subsection{Evaluation of Heuristic Range Filters}\label{ssec:exp-heuristic}

\begin{figure*}
    \centering
    \includegraphics[]{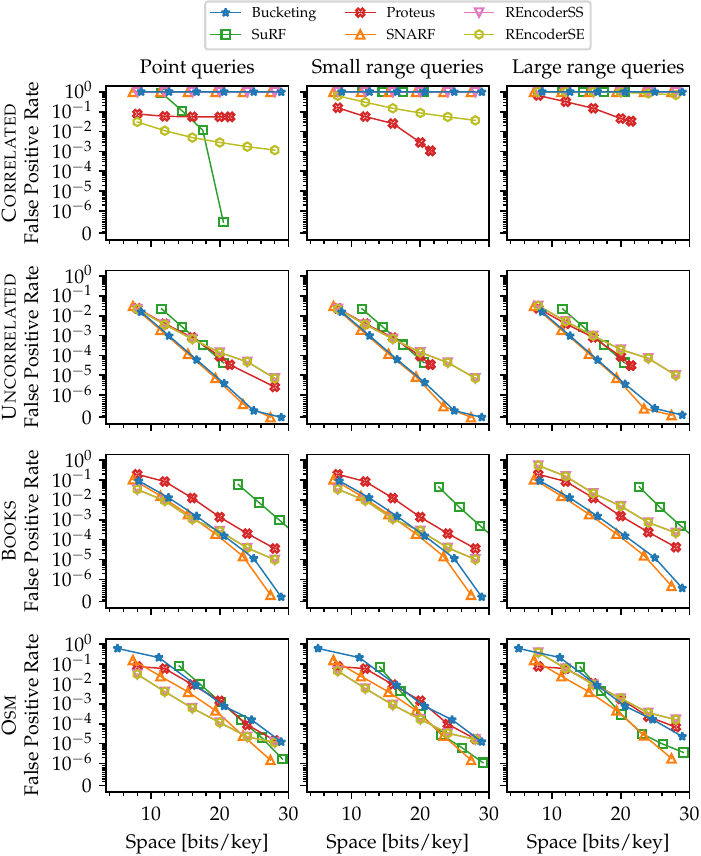}
    \begin{minipage}[b]{0.29\textwidth}
    \begin{tabular}{rl}
   \toprule
   Range filter & Avg ns/query \\
   \midrule
      Bucketing &                       67 (1.0$\times$) \\
      SuRF &                    275 (4.1$\times$) \\
      SNARF &                    576 (8.6$\times$) \\
      REncoderSS &                   732 (10.93$\times$) \\
      REncoderSE &                  4074 (60.81$\times$) \\
      Proteus &                 28638 (427.43$\times$) \\
   \bottomrule
   \vspace{.7em}\\
   \toprule
        Bucketing &                      70 (1.0$\times$) \\
        Proteus &                    227 (3.24$\times$) \\
        SuRF &                    228 (3.26$\times$) \\
        REncoderSE &                    351 (5.01$\times$) \\
        REncoderSS &                    352 (5.03$\times$) \\
        SNARF &                   895 (12.79$\times$) \\
   \bottomrule
   \vspace{.7em}\\
   \toprule
    Bucketing &                     277 (1.0$\times$) \\
    SuRF &                    423 (1.53$\times$) \\
    REncoderSS &                     527 (1.9$\times$) \\
    REncoderSE &                    528 (1.91$\times$) \\
    Proteus &                    573 (2.07$\times$) \\
    SNARF &                   1817 (6.56$\times$) \\
   \bottomrule
   \vspace{0.7em}\\
   \toprule
      Bucketing &                     456 (1.0$\times$) \\
        REncoderSE &                    466 (1.02$\times$) \\
        REncoderSS &                    467 (1.02$\times$) \\
       SuRF &                    567 (1.24$\times$) \\
      Proteus &                    719 (1.58$\times$) \\
      SNARF &                   2908 (6.38$\times$) \\
   \bottomrule
   \vspace{1.4em}\\
   \end{tabular}

    \end{minipage}
    \caption{Comparison among heuristic range filters. In the first row, only \proteus and \rencoderse provide some range query filtering (albeit unsatisfactorily, as discussed in Section~\ref{ssec:exp-robustness}) because they are auto-tuned on the correlated query workload. In the other rows, a simple solution like Bucketing provides very close or better FPR, and much better query time than all the other heuristic range filters. We remark that, unlike the other range filters, \snarf suffers from false negatives (see Footnote~\ref{foot:snarf_bug}).}
    \label{fig:fpr_test_heuristics}
\end{figure*}

Figure~\ref{fig:fpr_test_heuristics} shows the results of our experiments with heuristic range filters.
Each column of the plot corresponds to a query range size (point, small and large), and each row corresponds to a dataset (the first two rows are the \textsc{Correlated} and \textsc{Uncorrelated} query workloads we experiment on \textsc{Uniform} data, and the other two rows correspond to the \textsc{Books} and \textsc{Osm} datasets).
At the right of each row, we show a table with the query time of each range filter, averaged over the various space configurations and query range sizes, and next to each query time we show its ratio with respect to the fastest range filter.

In \textsc{Correlated}, in line with the experiment of Section~\ref{ssec:exp-robustness}, heuristic range filters provide no filtering (\snarf, \rencoderss, and \surf, whose performance on point queries is commented in Section~\ref{ssec:exp-robustness}) or little filtering (\proteus and \rencoderse). These last two filters are actually advantaged by being auto-tuned on the query workload, which might not be realistic in some applications due to rapidly-changing workloads or due to the additional space needed by query logs (which we did not account in their space usage).

For the other datasets, we notice that the filtering effectiveness of Bucketing essentially matches (on \textsc{Uncorrelated} and \textsc{Books}) or is very close (\textsc{Osm}) to the one of the best-performing heuristic range filter that is typically either \snarf (which, however, suffers from false negatives, see Footnote~\ref{foot:snarf_bug}) or \rencoderse/SS, while simultaneously providing up to 13$\times$ faster queries than \snarf and up to 5$\times$ faster queries than \rencoderse/SS. Moreover, Bucketing provides the best construction times, as we will show in Section~\ref{ssec:exp-construction}.

\subsection{Evaluation of Robust Range Filters}

\begin{figure*}
    \centering
    \includegraphics[]{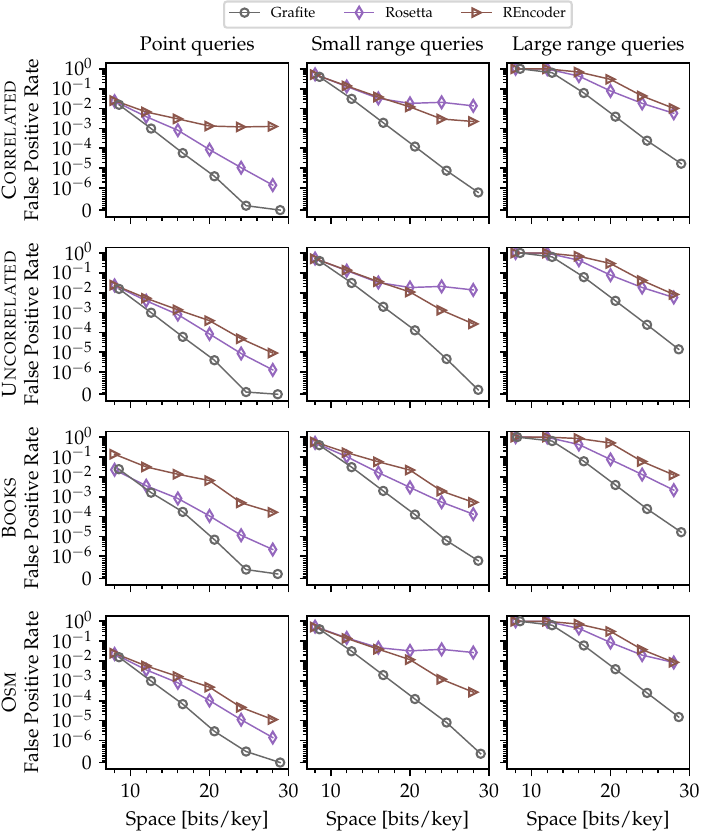}
    \begin{minipage}[b]{0.29\textwidth}
    \begin{tabular}{rl}
\toprule
Range filter & Avg ns/query \\
\midrule
   Grafite &                  305 (1.00$\times$) \\
  REncoder &                 2905 (9.52$\times$) \\
   Rosetta &               24902 (81.65$\times$) \\
\bottomrule
\vspace{4em}\\
\toprule
   Grafite &                  270 (1.00$\times$) \\
  REncoder &                3001 (11.11$\times$) \\
   Rosetta &               24913 (92.27$\times$) \\
\bottomrule
\vspace{4.4em}\\
\toprule
   Grafite &                   274 (1.00$\times$) \\
  REncoder &                  2410 (8.80$\times$) \\
   Rosetta &                24716 (90.20$\times$) \\
\bottomrule
\vspace{4.4em}\\
\toprule
   Grafite &                  276 (1.00$\times$) \\
  REncoder &                2845 (10.31$\times$) \\
   Rosetta &               24853 (90.05$\times$) \\
\bottomrule
\vspace{3em}\\
\end{tabular}

    \end{minipage}
    \caption{Grafite dominates all other robust range filters by providing up to 5 orders of magnitude better FPR and up to 92$\times$ faster queries. These substantial improvements, coupled with its performance guarantees (Corollary~\ref{cor:grafite}), make Grafite the range filter of choice in applications handling a variety of data distributions and query workloads, even adversarial ones.}
    \label{fig:fpr_test_bounded}
\end{figure*}

We now experiment with robust range filters, namely Grafite, \rosetta, and \rencoder (note this last one is slightly less robust in the case of small range queries, as discussed in Section~\ref{ssec:exp-robustness}).

\looseness=-1
As Figure~\ref{fig:fpr_test_bounded} shows, in all datasets and query range sizes, Grafite dominates  \rosetta and \rencoder both in terms of FPR and query time.
In particular, in terms of FPR, Grafite is up to 4 orders of magnitude more effective than \rencoder, and up to 5 orders of magnitude more effective than \rosetta.
In terms of query time, Grafite is 9.5--11.1$\times$ faster than \rencoder, and 81.7--92.3$\times$ faster than \rosetta.
Besides, we observe that Grafite has the most predictable FPR across all combinations of datasets, query workloads, and range sizes.

This consistent and substantial improvement of the state of the art corroborates the theoretical advantage of Grafite over prior solutions (Section~\ref{sec:theoretical}), and demonstrates its potential to become the range filter of choice in applications handling a variety of data distributions and query workloads (even adversarial ones).

\subsection{Performance on Non-Empty Queries}

\begin{figure*}[t]
    \centering
    \includegraphics[width=\linewidth]{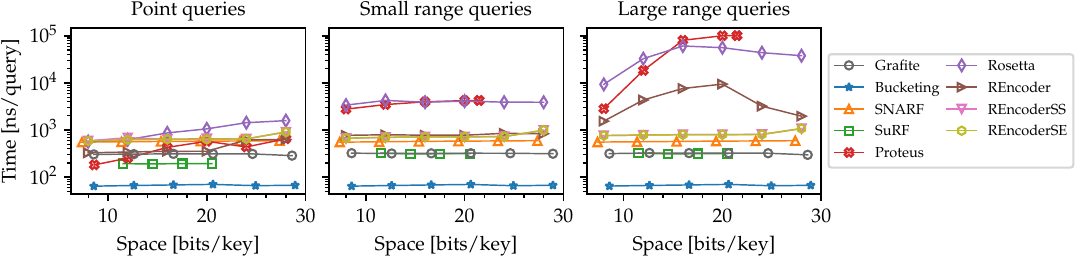}
    \caption{The query time of range filters can vary a lot also in the case of non-empty queries, as shown in these plots with a logarithmic time axis.  Grafite and Bucketing provide the best query times among robust and heuristic range filters, respectively.}
    \label{fig:true_queries_test}
\end{figure*}

We now experiment with queries that intersect the input dataset to show their impact on the query time.
We use the \textsc{Uniform} dataset and create a query range $[x, x+L-1]$ by first picking a key $k$ randomly from the dataset, and then picking the left endpoint $x$ randomly in $[k - L + 1, k]$. 

Figure~\ref{fig:true_queries_test} shows the results: among heuristic range filters, Bucketing provides up to 3 orders of magnitude faster queries than the others; among robust range filters, Grafite provides the fastest queries, up to 1 order of magnitude faster than \rencoder and up to 2 orders of magnitude faster than \rosetta.

A remark is necessary at this point. Even though filters are typically used in applications to prevent unnecessary (due to empty queries) network or disk accesses, they also increase CPU usage (regardless of the actual emptiness of the queried range).
In some cases, high CPU usage might not compensate for the reduction in access frequency to a slow resource, thus making the choice of a query-efficient range filter preferable, even if it has a higher FPR.
For example, \rosetta and \proteus in Figure~\ref{fig:true_queries_test} take up to 61.2 and 101.5 microseconds per query, respectively, which is comparable to the access latency of an SSD.
In other cases, the opposite might be true, i.e. the cost of accessing a slow resource might be too high to be able to afford a range filter with a lower FPR but better CPU usage; thus the choice of which range filter to use ultimately depends on the specific application.

\subsection{Construction Efficiency}\label{ssec:exp-construction}

\begin{figure}[t]
    \centering
    \includegraphics[width=\linewidth]{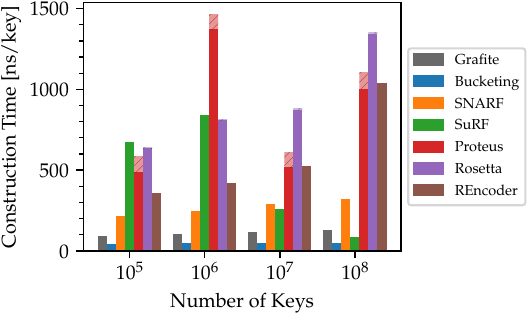}
    \caption{Grafite has the best construction time among robust range filters (\rosetta and \rencoder). Bucketing has the best construction time among heuristic range filters.}
    \label{fig:constr_time_test}
\end{figure}

Figure~\ref{fig:constr_time_test} shows the construction time of the various range filters as the number of keys increases from $10^5$ to $10^8$.
We use the \textsc{Uniform} dataset (other datasets do not change our conclusions) and average the construction time over the different space budgets.
We do not show \rencoderse and SS because their construction time is identical to that of \rencoder.
For both \rosetta and \proteus, the plot shows with a light colour the impact of the tuning process, which was evaluated with an \textsc{Uncorrelated} query workload of $n/10$ small range queries.

Among heuristic range filters, Bucketing is the fastest to construct, from 1.8 to 30.2$\times$ faster than the others. In particular, compared to its closest competitors in terms of FPR (see Section~\ref{ssec:exp-heuristic}), Bucketing is 8.6--23.9$\times$ faster to construct than \rencoderse and 4.9--6.5$\times$ faster than \snarf.

Among robust range filters, Grafite is the fastest to construct, 6.7--10.3$\times$ faster than \rosetta and 3.8--7.9$\times$ faster than \rencoder.

Furthermore, the construction time of both Grafite and Bucketing is linear for increasing values of $n$ (note the plot shows the construction time per key), which makes them highly scalable.

Actually, Grafite could achieve an even better construction time by using other sorting algorithms~\cite{Axtmann2022sorting} in place of the \texttt{spreadsort} algorithm we are using, or by enabling multi-threaded sorting (not shown in the plots for fairness with the single-threaded competitors).
For example, by using the parallel \texttt{block\_indirect\_sort} algorithm from the Boost library, the construction time for 200M keys with just 2 threads is reduced from 28.0 to 18.8 s (a 1.5$\times$ speedup), with 4 threads to 15.8 s (1.8$\times$ speedup), with 8 threads to 14.0 s (2.0$\times$ speedup)

\subsection{Discussion and Recommendations}\label{ssec:recommendations}

We now summarise our experimental findings by providing some guidance on which range filter to adopt for an application.

First and foremost, one should determine whether guarantees on the filtering effectiveness and query performance are needed regardless of the input data and future queries. If the answer is affirmative, then range filters providing a bounded FPR for a given space budget (and vice versa) are the choice, and among them Grafite is the best option.

For example, if future queries are correlated (i.e. close) to the input keys, the existing heuristic range filters provide little to no filtering, thus impacting the overall performance of the system (and possibly cloud costs) due to the network or disk accesses the filters are deployed to prevent. 
Correlated queries are common in practice~\cite{luoRosettaRobustSpacetime2020}, and malicious users can artificially issue them with just the knowledge of (a subset of) the keys.
In these cases, Grafite is again the best option since it is unaffected by correlated queries.

If the application has no or infrequent correlated queries, and the query distribution does not change after the range filter is evaluated on a query sample and deployed, we recommend considering also Bucketing, \proteus, \rencoderss (and possibly \snarf, but refer to Footnote~\ref{foot:snarf_bug}), which could provide better filtering effectiveness than Grafite.
For example, \proteus can auto-tune itself and obtain a good FPR (see the rightmost plot in Figure~\ref{fig:corr_test}). \rencoderss can offer a good FPR without any auto-tuning in some cases with small range queries (see last two rows of Figure~\ref{fig:fpr_test_heuristics}). Bucketing  always offered the best query and construction times in our experiments, and very good FPR in many cases (see Figure~\ref{fig:fpr_test_heuristics}).
Other than the FPR, query and construction times, deciding which range filter to adopt in a real application should consider factors like the cost of a false positive (e.g., in terms of latency or cloud costs) and the frequency of queries (which impact on the CPU usage). Thus the best choice ultimately depends on the peculiarities of the  application.

\section{Conclusion}\label{sec:conclusion}

We introduced Grafite, a range filter that solves the lack of robustness in current practical solutions by providing strong theoretical guarantees on the false positive probability, optimal space usage, and very efficient and effective performance across many datasets, query workloads, and range sizes.
We also introduced Bucketing, which simplifies the design of existing heuristic range filters while empirically providing very close or better filtering effectiveness, and much faster query and construction times, thus possibly resulting in a simple substitute for them.

For future work, we mention a more in-depth study of the Bucketing approach, which could be made workload-aware (e.g. by creating larger buckets for key ranges that are queried less frequently), or combined with Grafite.
It is also worth engineering and experimenting with an extension of Grafite to string keys, for example by treating strings as integers and choosing $r$ as a power of two, say $r=2^k$ for some $k>0$, so that the hash function~\eqref{eq:hash} can be efficiently implemented via bitwise and arithmetic operations as $h(x) = (q(x \,\mathtt{\gg}\, k) + x) \,\mathtt{\&}\, (r-1)$, where $q$ could be chosen to be a practical hash function for strings like xxHash. Another open problem is to support the insertion of new keys in Grafite and Bucketing, for which dynamic Elias-Fano representations could help~\cite{pibiriDynamicEliasfanoRepresentation2017}.
Finally, we mention again that Grafite can easily return an approximate count of the keys that intersect the given query range without any change in its space or query time complexity, thus potentially being a practical and efficient solution for this other interesting problem too~\cite{Alstrup2001reporting}.

\begin{acks}
\looseness=-1
We thank Mayank Goswami and Rasmus Pagh for  clarifications on~\cite{goswamiApproximateRangeEmptiness2014}, Subarna Chatterjee for clarifications on~\cite{luoRosettaRobustSpacetime2020}, and \mbox{Hans-Peter} Lehmann for noticing a special case in our query algorithm.

This work was supported by the European Union -- Horizon 2020 Program under the scheme ``INFRAIA-01-2018-2019 -- Integrating Activities for Advanced Communities'', Grant Agreement n. 871042, ``SoBigData++: European Integrated Infrastructure for Social Mining and Big Data Analytics'' (http://www.sobigdata.eu), by the NextGenerationEU -- National Recovery and Resilience Plan (Piano Nazionale di Ripresa e Resilienza, PNRR) -- Project: ``SoBigData.it - Strengthening the Italian RI for Social Mining and Big Data Analytics'' -- Prot. IR0000013 -- Avviso n. 3264 del 28/12/2021, by the spoke ``FutureHPC \& BigData'' of the ICSC -- Centro Nazionale di Ricerca in High-Performance Computing, Big Data and Quantum Computing funded by European Union -- NextGenerationEU -- PNRR, by the  Italian Ministry of University and Research ``Progetti di Ri\-le\-van\-te In\-te\-res\-se Na\-zio\-nale'' project: ``Multicriteria data structures and algorithms'' (grant n. 2017WR7SHH).
\end{acks}
\balance

\bibliographystyle{ACM-Reference-Format}
\bibliography{bibliography}

\end{document}